\documentclass[a4paper,11pt]{article}
\usepackage[pdftex]{color,graphicx}
\usepackage{pdfpages}
\usepackage{amsmath,amssymb}

\pdfoutput=1

\newcommand{\assign}{:=}
\newcommand{\tmem}[1]{{\em #1\/}}
\newcommand{\tmname}[1]{\textsc{#1}}
\newcommand{\tmop}[1]{\ensuremath{\operatorname{#1}}}
\newcommand{\tmsamp}[1]{\textsf{#1}}
\newcommand{\tmstrong}[1]{\textbf{#1}}
\newcommand{\tmtextit}[1]{{\itshape{#1}}}
\newcommand{\tmtexttt}[1]{{\ttfamily{#1}}}

\begin{document}

  \thispagestyle{empty}



\hspace*{-.5cm}
\begin{minipage}{18cm}

  \noindent\includegraphics[height=2.1cm]{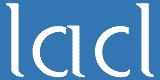}\hfill
  \includegraphics[height=2.1cm]{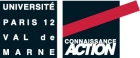}

  \vspace*{0.5cm}

  \hrule

  \vspace*{3cm}

  \begin{center}
    {\LARGE\bf LoopW Technical Reference v0.3}
  \end{center}

  \vspace*{2cm}

  \begin{center}
    {\bf Emmanuel Polonowski}
  \end{center}

  \vspace*{5cm}

  \begin{center}
    {\it December\ \ 2009}\\[1ex]
    {\large\bf TR--LACL--2009--8}
  \end{center}

  \vspace*{6cm}

  \hrule

  {\begin{center}
      \textbf{Laboratoire d'Algorithmique, Complexit\'e et Logique (LACL)}\\
	     {\small{\bf  D\'epartement d'Informatique}\\
	       {\bf Universit\'e Paris~Est -- Val~de~Marne, Facult\'e des Science et
		 Technologie}\\
	       61, Avenue du G\'en\'eral de Gaulle, 94010 Cr\'eteil cedex, France\\
	       Tel.: (33)(1) 45 17 16 47,\ \  Fax: (33)(1) 45 17 66 01}
  \end{center}}

\end{minipage}

  \clearpage

  \thispagestyle{empty}

  \vspace*{3cm}

  \begin{center}
    Laboratory of Algorithmics, Complexity and Logic (LACL)\\
    University Paris 12 (Paris Est)\\[1ex]
    Technical Report {\bf TR--LACL--2009--8}\\[2ex]
    E.~Polonowski.\\ {\it LoopW Technical Reference v0.3}\\
  \end{center}

  \copyright\ \  E.~Polonowski,  December 2009.

  \clearpage

\setcounter{page}{1}

\title{LoopW\\
Technical Reference}\author{Emmanuel Polonowski}\date{}\maketitle

\begin{abstract}
  This document describes the implementation of the {\tmsamp{LoopW}} language,
  an imperative language with higher-order procedural variables and non-local
  jumps equiped with a program logic, in {\tmsamp{SML}}. It includes the user
  manual along with some implementation notes and many examples of certified
  imperative programs. As a concluding example, we show the certification of
  an imperative program encoding {\tmsamp{shift/reset}} using
  {\tmsamp{callcc/throw}} and a global meta-continuation.
\end{abstract}

\section{Introduction}

This document describes the implementation in Standard ML of the
{\tmsamp{LoopW}} language~{\cite{CROLARD:2009:HAL-00422158:1,CP10}}, an
imperative language with higher-order procedural variables and non-local jumps
equiped with a program logic.

This implementation consists firstly in a type inference tool prototype
\tmtexttt{LoopW} that reads partially annotated imperative source code and
performs combinated type checking and type inference; and secondly in a proof
checking tool prototype \tmtexttt{LoopWc} that reads completely annotated
imperative source code and performs proof checking, translation into
completely annotated functional source code and proof checking of the
functional code.

The complete source code archive of those prototypes along with numerous small
examples of certified programs can be obtained at
\tmtexttt{http://lacl.univ-paris12.fr/polonowski/Develop/LoopW/loopw.html}.

Section~\ref{sect-manual} contains the user manual, describing the source
syntax, the usage of the tools \tmtexttt{LoopW} and \tmtexttt{LoopWc}, and a
small introduction on imperative program certification with examples.
Section~\ref{sect-notes} gives details about the current implementation, along
with small commented protions of Standard ML source code.
Section~\ref{sect-smldoc} presents the overview of the source code
documentation (generated using \tmtexttt{smldoc} -- from SML\# --
\tmtexttt{http://www.pllab.riec.tohoku.ac.jp/smlsharp/}) which is included in
the source code archive.

{\newpage}

\section{User Manual\label{sect-manual}}

\subsection{Source syntax}

\begin{center}
  \begin{tabular}{llcll}
    (terms) & \tmtexttt{t} & ::= & \tmtexttt{i} & {\tmem{term variable}}\\
    &  & $|$ & \tmtexttt{f(t, ..., t)} & {\tmem{function application}}\\
    &  & $|$ & \tmtexttt{0} & {\tmem{zero}}\\
    &  & $|$ & \tmtexttt{s(t)} & {\tmem{successor}}\\
    &  & $|$ & \tmtexttt{p(t)} & {\tmem{predecessor}}\\
    &  & $|$ & \tmtexttt{t + t} & {\tmem{addition}}\\
    &  & $|$ & \tmtexttt{t * t} & {\tmem{multiplication}}\\
    &  & $|$ & \tmtexttt{t - t} & {\tmem{substraction}}\\
    &  &  &  & \\
    &  &  &  & \\
    (types) & \tmtexttt{T} & ::= & \tmtexttt{\$} & {\tmem{absurd}}\\
    &  & $|$ & \tmtexttt{(t = t)} & {\tmem{equality over terms}}\\
    &  & $|$ & \tmtexttt{nat(t)} & {\tmem{nat predicate}}\\
    &  & $|$ & \tmtexttt{proc(\{i, ..., i\} in T, ..., T;} & {\tmem{procedure
    prototype}}\\
    &  &  & \tmtexttt{ \ \ \ \ \{i, ..., i\} out T, ..., T)} & \\
    &  & $|$ & \tmtexttt{\~{ }T} & {\tmem{negation}}\\
    &  &  &  & \\
    &  &  &  & \\
    (expressions) & \tmtexttt{e} & ::= & $\bar{q}$ & {\tmem{natural numbers
    constants}}\\
    &  & $|$ & \tmtexttt{X} & {\tmem{program variable}}\\
    &  & $|$ & \tmtexttt{*} & {\tmem{true value}}\\
    &  & $|$ & \tmtexttt{proc(\{i, ..., i\} in X:T, ..., X:T;} &
    {\tmem{anonymous procedure declaration}}\\
    &  &  & \tmtexttt{ \ \ \ \ \{i, ..., i\} out X:T, ..., X:T) \{ s \}} & \\
    &  &  &  & \\
    &  &  &  & \\
    (commands) & \tmtexttt{c} & ::= & \tmtexttt{\{ s \}X:T, ..., X:T} &
    {\tmem{annotated block}}\\
    &  & $|$ & \tmtexttt{for I := 0 until e \{ s \}X:T, ..., X:T} &
    {\tmem{annotated for loop}}\\
    &  & $|$ & \tmtexttt{X := e} & {\tmem{assignment}}\\
    &  & $|$ & \tmtexttt{inc(X)} & {\tmem{incrementation}}\\
    &  & $|$ & \tmtexttt{dec(X)} & {\tmem{decrementation}}\\
    &  & $|$ & \tmtexttt{e(e, ..., e; X, ..., X)} & {\tmem{procedure call}}\\
    &  & $|$ & \tmtexttt{X : \{ s \}X:T, ..., X:T} & {\tmem{labeled block}}\\
    &  & $|$ & \tmtexttt{jump(e, e, ..., e)X:T, ..., X:T} & {\tmem{jump to a
    label}}\\
    &  &  &  & \\
    &  &  &  & \\
    (sequences) & \tmtexttt{s} & ::= & $\varepsilon$ & {\tmem{empty
    sequence}}\\
    &  & $|$ & \tmtexttt{c; s} & {\tmem{command composition}}\\
    &  & $|$ & \tmtexttt{cst X = e; s} & {\tmem{constant declaration}}\\
    &  & $|$ & \tmtexttt{var X := e; s} & {\tmem{variable declaration}}
  \end{tabular}
\end{center}

{\newpage}

\subsection{Usage}

\begin{verbatim}
========== LoopW imperative proof inference and proof checking program ==========

This tool contains two executables, namely loopW and loopWc.

I. loopW: imperative proof inference.

The loopW inference tool reads its input in a file with extension ".loop" and tries to
infer a correct and complete proof derivation that corresponds to the given source code.

usage: loopW [-version] [-v123] [-uprint] [-print] [-pprint] [-form] [-tex] [-ott] file.loop
  -version print version and exit
  -v Verbosity level (1 : low, 2 : medium, 3 : high)
  -uprint Erase all type information and pretty print (file.cs)
  -print Pretty print (file.rev.loop)
  -pprint Complete infered proof pretty printing (file.proof)
  -form Display formula-like types instead of imperative types
  -tex Tex output
  -ott Ott/twelf complete infered proof pretty printing (file.loop.ott)

NOTES
- On Unix systems (including Mac OS X) you need to have mono (www.mono-project.com)
  installed and you can then run the command as:

  mono loopW.exe [-version] [-v123] [-uprint] [-print] [-pprint] [-form] [-tex] [-ott] file.loop

II. loopWc: imperative proof checking, functional translation and proof checking.

The loopWc proof checker and translater tool reads its input in a file with extension
".proof" (eventually written by the loopW tool (see above)). It first checks the imperative
proof derivation, then it translate it into a functional proof derivation and check it.

usage: loopWc [-version] [-v123] [-uprint] [-fprint] file.proof
  -version print version and exit
  -v Verbosity level (1 : low, 2 : medium, 3 : high)
  -uprint Erase all type information and pretty print (file.pcs)
  -fprint Functional translation (untyped) pretty print (file.sml)

NOTES
- On Unix systems (including Mac OS X) you need to have mono (www.mono-project.com)
  installed and you can then run the command as:

  mono loopWc.exe [-version] [-v123] [-uprint] [-fprint] file.proof

- The generated sml file can be tested with SML/NJ with the provided library file:

  sml lib.sml file.sml
\end{verbatim}

{\newpage}

\subsection{Program certification examples}

\subsubsection{Addition}

\begin{verbatim}
cst p_add = proc({x, y} in X:nat(x), Y:nat(y); out Z:nat(x + y)) {
    Z := X :> nat(x + 0);
    for i := 0 until Y {
        inc(Z);
    }Z:nat(x + i);
};
var N := *;
p_add(3, 5; N);
\end{verbatim}

To compile this example (in a file \tmtexttt{add.loop}), we use the command:
\tmtexttt{loopW add.loop}. The absence of output indicates that the
compilation was successfull. We find a new file \tmtexttt{add.typ} containing
the following text:

\begin{verbatim}
cst p_add = proc(in X, Y; out Z) {   -(X:nat(x), Y:nat(y))[Z:(0 = 0)]
    Z := X;                          |   [Z:nat((x + 0))]    by #1
    for i := 0 until Y {             |   -(i:nat(i))[Z:nat((x + i))]
        inc(Z);                      |   |   [Z:nat(s((x + i)))]
    }Z;                              |   [Z:nat((x + y))]    by #2
};                                   (p_add:proc({x, y} in nat(x), nat(y); out nat((x + y))))
var N := *;                          [N:(0 = 0)]
p_add(3, 5; N);                      [N:nat((3 + 5))]

1: |- (x = (x + 0))
2: |- (s((x + i)) = (x + s(i)))
\end{verbatim}

We find on the left part our source code (without type informations) and on
the right part the typing derivation. Remark that variables may change their
type: \tmtexttt{N} is declared and assigned with \tmtexttt{*} of type
\tmtexttt{0 = 0}, and after the procedure call, its type is \tmtexttt{nat(3 +
5)}.

The bottom part of the file contains the equalities used to certify the source
code. They must be checked (either manually or with a solver) in order to
guarantee the correctness of the code. Here, the equalities are
straightforward consequences of the usual axioms defining addition for peano
arithmetic.

In order to proof-check this program, we need to generate the intermediate
file \tmtexttt{add.proof}. We use the command \tmtexttt{loopW -pprint
add.loop}. We can then call the proof-checker: \tmtexttt{loopWc add.proof}.
Again, the absence of output indicates that the proof-checking was
successfull. We can also generate the functional program: \tmtexttt{loopWc
-fprint add.proof}. It gives us the following file \tmtexttt{add.sml}:

\begin{verbatim}
let val p_add (* : Forall x,y.((nat(x) & nat(y)) => (nat((x + y)))) *) =
  fn (X (* : nat(x) *), Y (* : nat(y) *)) =>
    let val Z (* : nat((x + 0)) *) =
          X (* : nat(x) *) (* :> {var_1/nat(var_1)}[x = (x + 0) by #1] *)
        val Z (* : nat((x + y)) *) =
          Rec (Y (* : nat(y) *), Z (* : nat((x + 0)) *),
               fn i (* : nat(i) *) => fn (Z (* : nat((x + i)) *)) =>
                 let val Z (* : nat(s((x + i))) *) = Succ(Z (* : nat((x + i)) *))
                 in Z (* : nat(s((x + i))) *)
                      (* :> {var_2/(nat(var_2))}[s((x + i)) = (x + s(i)) by #2] *) end )
        in Z (* : nat((x + y)) *) end
      val N (* : 0 = 0 *) = ()
      val N (* : nat((3 + 5)) *) =
        p_add (* : Forall x,y.((nat(x) & nat(y)) => (nat((x + y)))) *)((3, 5)) (* [3, 5] *)
in () end

(* 1:  |- x = (x + 0) *)
(* 2:  |- s((x + i)) = (x + s(i)) *)
\end{verbatim}

If we erase all the comments containing the typing informations, we get:

\begin{verbatim}
let val p_add = fn (X, Y) =>
    let val Z = X
        val Z = Rec (Y, Z, fn i => fn (Z) =>
                let val Z = Succ(Z)
                in Z end)
        in Z end
      val N = ()
      val N = p_add(3, 5)
in () end

(* 1:  |- x = (x + 0) *)
(* 2:  |- s((x + i)) = (x + s(i)) *)
\end{verbatim}

It clearly corresponds to a valid definition of addition using a recursion
operator.

\subsubsection{Ackermann}

\begin{verbatim}
cst ack = proc ({m,n} in M : nat(m), N : nat(n); out Z : nat(a(m,n))) {
    var G := proc ({y} in Y : nat(y); out P : nat(a(0,y))) {
        P := Y;
        inc(P);
    };
    for i := 0 until M {
        cst H = G;
        G := proc ({y} in Y : nat(y); out P : nat(a(s(i),y))) {
            P := 2 :>  nat(a(s(i),0));
            for j := 0 until Y {
                H(P; P);
            }P:nat(a(s(i),j));
        };
    }G:proc ({y} in nat(y); {} out nat(a(i,y)));
    G(N; Z);
};
\end{verbatim}

\subsubsection{Negation}

Using the negation, we can use the full power of classical logic to prove
theorems. For instance, the following program is a proof of $\neg (\forall x.
\tmop{nat} (x) \wedge \neg F) \Rightarrow \exists x. \tmop{nat} (x) \wedge F$
which is not provable in intuitionnistic logic.

\begin{verbatim}
cst dblNegElim = proc (in K : ~~F; out Z : F) {
    K2: {
        jump(K,K2)Z:F;
    }Z:F;
};

cst notAllNot_implies_Exist = proc (in H : ~proc({x} in nat(x); out ~F);
    			   	    out C : proc(; {x} out nat(x), F)) {
    K: {
	cst P = proc ({x} in N : nat(x); out Z : ~F) {
            K2: {
                dblNegElim(K2; Z);
                cst V = Z;
                cst Q = proc(; {x} out M : nat(x), Y : F) {
                    M := N;
                    Y := V;
                };
                jump(K, Q)Z:~F;
            }Z:~F;
        };
        jump(H, P)C:proc(; {x} out nat(x), F);
    }C:proc(; {x} out nat(x), F);
};
\end{verbatim}

\subsubsection{Shift/reset}

\begin{verbatim}
cst shift = proc (in p : proc(in proc({n} in nat(n), ~A; out nat(F32(n)), ~A),
                                 ~proc ({n} in nat(n), ~A; out nat(F32(n)), ~A);
                              out proc ({n} in nat(n), ~A; out nat(F32(n)), ~A),
                                  ~proc ({n} in nat(n), ~A; out nat(F32(n)), ~A)),
                     mk2 : ~proc ({n} in nat(n), ~A; out nat(F32(n)), ~A);
                  {u} out r : nat(u), mk : ~nat(F32(u))) {
    mk := mk2;
    cst reset2 = proc ({x} in p : proc(in ~nat(F32(x)); out H,~H), mk2 : ~A;
                       out r : nat(F32(x)), mk : ~A) {
    mk := mk2;

    k: {
        cst m = mk;

        mk := proc (in r : nat(F32(x)); out Z : $) {
            jump(k, r, m)Z:$;
        };

        var y := *;
        p(mk; y, mk);
        jump(mk, y)r:nat(F32(x)), mk:~A;
    }r:nat(F32(x)), mk:~A;

};

    k: {

        cst q = proc ({x} in v : nat(x), mk2 : ~A;
                      out r : nat(F32(x)), mk : ~A) {
            mk := mk2;

            cst anonym = proc(in mk2 : ~nat(F32(x)); out z : H, mk : ~H) {
                mk := mk2;
                jump(k, v, mk)z:H,mk:~H;
            };
            reset2(anonym, mk; r, mk);
        };

        var y := *;
        p(q, mk; y, mk);
        jump(mk, y)r:nat(0), mk:~nat(F32(0));
    }{u}r:nat(u), mk:~nat(F32(u));
};

cst reset = proc (in p : proc(in ~proc({n} in nat(n), ~A; out nat(F32(n)), ~A);
                              {v} out nat(v), ~nat(v)),
                     mk2 : ~A;
                  out r : proc({n} in nat(n), ~A; out nat(F32(n)), ~A),
                      mk : ~A) {
    mk := mk2;

    k: {
        cst m = mk;

        mk := proc (in r : proc({n} in nat(n), ~A; out nat(F32(n)), ~A); out Z : $) {
            jump(k, r, m)Z:$;
        };

        var y := *;
        p(mk; y, mk);
        jump(mk, y)r:proc({n} in nat(n), ~A; out nat(F32(n)), ~A), mk:~A;
    }r:proc({n} in nat(n), ~A; out nat(F32(n)), ~A), mk:~A;

};

cst a = proc (in mk2 : ~A; out z : nat(3+2), mk : ~A) {
    cst p_add = proc ({x,y} in X : nat(x), Y : nat(y), mk2 : ~A;
                      out Z : nat(x + y), mk : ~A) {
        mk := mk2;
        Z := X :> nat(x + 0);
        for i := 0 until Y {
            inc(Z);
        }Z:nat(x + i);
    };
    cst q = proc(in mk2 : ~proc({n} in nat(n), ~A; out nat(F32(n)), ~A);
                {v} out r : nat(v), mk : ~nat(v)) {
        mk := mk2;
        cst p = proc(in f : proc ({n} in nat(n), ~A; out nat(F32(n)), ~A),
                        mk2 : ~proc ({n} in nat(n), ~A; out nat(F32(n)), ~A);
                     out h : proc ({n} in nat(n), ~A; out nat(F32(n)), ~A),
                         mk : ~proc ({n} in nat(n), ~A; out nat(F32(n)), ~A)) {
            mk := mk2;
            h := f;
        };
        var b := *;
        shift(p, mk; b, mk);
        r := 3 :> nat(F32(0));
        for i := 0 until b {
            r := 2 :> nat(F32(s(i)));
        }r:nat(F32(i));
    };
    var mk := mk2;
    var g := *;
    reset(q, mk; g, mk);
    var x := *;
    g(0, mk; x, mk);
    var y := *;
    g(1, mk; y, mk);
    p_add(x :> nat(3), y :> nat(2), mk; z, mk);
};
\end{verbatim}

{\newpage}

\section{Implementation Notes\label{sect-notes}}

\subsection{ProofChecker source code}

Here follows the formal definition of the imperative dependent type system.

{\small
\begin{eqnarray*}
    & \frac{\begin{array}{c}
      x : \tau \in \Gamma ; \Omega
    \end{array}}{\begin{array}{c}
      \Gamma ; \Omega \vdash x : \tau
    \end{array}} & \text{({\tmname{t.env}})}\\
    & \frac{\begin{array}{c}
      
    \end{array}}{\begin{array}{c}
      \Gamma ; \Omega \vdash \bar{q} : \text{{\tmstrong{nat}}} (
      \text{{\tmstrong{s}}}^q ( \text{{\tmstrong{0}}}))
    \end{array}} & \text{({\tmname{t.num}})}\\
    &  & \\
    & \frac{\begin{array}{c}
      \vdash_{\mathcal{E}} n = m
    \end{array}}{\begin{array}{c}
      \Gamma ; \Omega \vdash \ast : n = m
    \end{array}} & \text{({\tmname{t.equal}})}\\
    &  & \\
    & \frac{\begin{array}{c}
      \Gamma ; \Omega \vdash e : \tau [n / i] \hspace{2em} \Gamma ; \Omega
      \vdash e : n = m
    \end{array}}{\begin{array}{c}
      \Gamma ; \Omega \vdash e : \tau [m / i]
    \end{array}} & \text{({\tmname{t.subst-i}})}\\
    &  & \\
    & \frac{\begin{array}{c}
      \Gamma ; \Omega \vdash s \vartriangleright \Omega [n / i] \hspace{2em}
      \Gamma ; \Omega \vdash e : n = m
    \end{array}}{\begin{array}{c}
      \Gamma ; \Omega \vdash s \vartriangleright \Omega [m / i]
    \end{array}} & \text{({\tmname{t.subst-ii}})}\\
    &  & \\
    & \frac{\begin{array}{c}
      
    \end{array}}{\begin{array}{c}
      \Gamma ; \Omega \vdash \varepsilon \vartriangleright \Omega
    \end{array}} & \text{({\tmname{t.empty}})}\\
    &  & \\
    & \frac{\begin{array}{c}
      \Gamma ; \Omega \vdash e : \tau \hspace{2em} \Gamma, y : \tau ; \Omega
      \vdash s \vartriangleright \Omega'
    \end{array}}{\begin{array}{c}
      \Gamma ; \Omega \vdash \text{{\tmstrong{cst}}} \text{ } y = e ; \text{ }
      s \; \vartriangleright \Omega'
    \end{array}} & \text{({\tmname{t.cst}})}\\
    &  & \\
    & \frac{\begin{array}{c}
      \Gamma ; \Omega \vdash e : \tau \hspace{2em} \Gamma ; \Omega, y : \tau
      \vdash s \vartriangleright \Omega', y : \tau'
    \end{array}}{\begin{array}{c}
      \Gamma ; \Omega \vdash \text{{\tmstrong{var}}} \text{ } y \assign e ;
      \text{ } s \; \vartriangleright \Omega'
    \end{array}} & \text{({\tmname{t.var}})}\\
    &  & \\
    & \frac{\begin{array}{c}
      \Gamma ; \vec{x} : \vec{\sigma} \vdash s \vartriangleright \vec{x} :
      \vec{\tau} \hspace{2em} \Gamma ; \Omega, \vec{x} : \vec{\tau} \vdash s'
      \vartriangleright \Omega', \vec{x} : \vec{\tau}'
    \end{array}}{\begin{array}{c}
      \Gamma ; \Omega, \vec{x} : \vec{\sigma} \vdash \{s\}_{\vec{x}} ; s'
      \vartriangleright \Omega', \vec{x} : \vec{\tau}'
    \end{array}} & \text{({\tmname{t.block}})}\\
    &  & \\
    & \frac{\begin{array}{c}
      \Gamma ; \Omega, y : \text{{\tmstrong{nat}}} ( \text{{\tmstrong{s}}}
      (n)) \vdash s \vartriangleright \Omega', y : \tau
    \end{array}}{\begin{array}{c}
      \Gamma ; \Omega, y : \text{{\tmstrong{nat}}} (n) \vdash
      \text{{\tmstrong{inc}}} (y) ; s \vartriangleright \Omega', y : \tau
    \end{array}} & \text{({\tmname{t.inc}})}\\
    &  & \\
    & \frac{\begin{array}{c}
      \Gamma ; \Omega, y : \text{{\tmstrong{nat}}} ( \text{{\tmstrong{p}}}
      (n)) \vdash s \vartriangleright \Omega', y : \tau
    \end{array}}{\begin{array}{c}
      \Gamma ; \Omega, y : \text{{\tmstrong{nat}}} (n) \vdash
      \text{{\tmstrong{dec}}} (y) ; s \vartriangleright \Omega', y : \tau
    \end{array}} & \text{({\tmname{t.dec}})}\\
    &  & \\
    & \frac{\begin{array}{c}
      \Gamma ; \Omega, y : \sigma \vdash e : \tau \hspace{2em} \Gamma ;
      \Omega, y : \tau \vdash s \vartriangleright \Omega', y : \tau'
    \end{array}}{\begin{array}{c}
      \Gamma ; \Omega, y : \sigma \vdash y \assign e ; s \vartriangleright
      \Omega', y : \tau'
    \end{array}} & \text{({\tmname{t.assign}})}\\
    &  & \\
    & \frac{\begin{array}{c}
      \Gamma ; \Omega, \vec{x} : \vec{\sigma} [ \text{{\tmstrong{0}}} / i]
      \vdash e : \text{{\tmstrong{nat}}} (n) \hspace{1em} \Gamma, y :
      \text{{\tmstrong{nat}}} (i) ; \vec{x} : \vec{\sigma} \vdash s
      \vartriangleright \vec{x} : \vec{\sigma} [ \text{{\tmstrong{s}}} (i) /
      i] \hspace{1em} \Gamma ; \Omega, \vec{x} : \vec{\sigma} [n / i] \vdash
      s' \vartriangleright \Omega', \vec{x} : \vec{\sigma}'
    \end{array}}{\begin{array}{c}
      \Gamma ; \Omega, \vec{x} : \vec{\sigma} [ \text{{\tmstrong{0}}} / i]
      \vdash \text{{\tmstrong{for}}} \text{ } y \assign 0 \text{ }
      \text{{\tmstrong{until}}} \text{ } e \text{ } \{s\}_{\vec{x}} ; s'
      \vartriangleright \Omega', \vec{x} : \vec{\sigma}'
    \end{array}} & \text{({\tmname{t.for}})}^{\star}\\
    &  & \\
    & \frac{\begin{array}{c}
      \vec{z} \neq \emptyset \hspace{2em} \Gamma, \vec{y} : \vec{\sigma} ;
      \vec{z} : \vec{\top} \vdash s \vartriangleright \vec{z} : \vec{\tau} [
      \vec{u} / \vec{\jmath}]
    \end{array}}{\begin{array}{c}
      \Gamma ; \Omega \vdash \text{{\tmstrong{proc}}} \text{ } (
      \text{{\tmstrong{in}}} \text{ } \vec{y} ; \text{{\tmstrong{out}}} \text{
      } \vec{z})\{s\}_{\vec{z}} : \text{{\tmstrong{proc}}} \text{ } (\{
      \vec{\imath} \} \text{{\tmstrong{in}}} \text{ } \vec{\sigma} ; \{
      \vec{\jmath} \} \text{{\tmstrong{out}}} \text{ } \vec{\tau})
    \end{array}} & \text{({\tmname{t.proc}})}^{\ast}\\
    &  & \\
    & \frac{\begin{array}{c}
      \Gamma ; \Omega, \vec{r} : \vec{\omega} \vdash p :
      \text{{\tmstrong{proc}}} \text{ } (\{ \vec{\imath} \}
      \text{{\tmstrong{in}}} \text{ } \vec{\tau} ; \{ \vec{\jmath} \}
      \text{{\tmstrong{out}}} \text{ } \vec{\sigma}) \hspace{1em} \Gamma ;
      \Omega, \vec{r} : \vec{\omega} \vdash \vec{e} : \vec{\tau} [ \vec{u} /
      \vec{\imath}] \hspace{1em} \Gamma ; \Omega, \vec{r} : \vec{\sigma} [
      \vec{u} / \vec{\imath}] \vdash s \vartriangleright \Omega', \vec{r} :
      \vec{\sigma}'
    \end{array}}{\begin{array}{c}
      \Gamma ; \Omega, \vec{r} : \vec{\omega} \vdash p ( \vec{e} ; \vec{r}) ;
      s \vartriangleright \Omega', \vec{r} : \vec{\sigma}'
    \end{array}} & \text{({\tmname{t.call}})}^{\ast}\\
    &  & \\
    &  & \\
    & \text{$^{\ast}$where $\vec{\imath} \notin
    \mathcal{F}\mathcal{V}(\Gamma)$ in ({\tmname{t.proc}}), $\vec{\jmath}
    \notin \mathcal{F}\mathcal{V}(\Gamma, \Omega, \Omega', \vec{\sigma}')$ in
    ({\tmname{t.call}}) and $i \notin \mathcal{F}\mathcal{V}(\Gamma)$ in
    ({\tmname{t.for}})} & \\
    &  & 
  \end{eqnarray*}
}

Its implementation is given below. The main difference is that in the current
implementation, type variables are allowed (even if no unification is
performed on them).

\begin{verbatim}
(**
 * check that the given sequence represents a correct proof.
 *
 * @params gamma omega (j, omega') reg seq
 * @param gamma the constant proof checking environment.
 * @param omega the variable proof checking environment.
 * @param j the out type existential quantified variables.
 * @param omega' the out type.
 * @param reg the region.
 * @param seq the sequence to check.
 *
 * @return true if and only if seq is a correct proof.
 *)
fun checkSequence gamma omega (j, omega') reg seq =
  let val (iseq, (k, delta)) = seq
  in (* Checks whether the out type is equal to the sequence type annotation *)
    ((envEquals delta omega') andalso (Utils.listEquals k j)
     andalso (checkInSequence gamma omega (k, delta) reg iseq))
         (* Checks the internal sequence *)
    orelse typeFailureSeq gamma omega (j, omega') seq reg
      "ProofChecker.checkSequence: out type and annotation mismatch"
  end

and checkInSequence gamma omega (j, omega') reg = fn
  (* Rule T.EMPTY *)
    Empty (subst) =>
    (* We apply the substitution to build the instantiated type *)
    let val out_subst = substituteAllList subst omega'
      (* We restrict omega to omega' *)
      val delta = List.filter (fn (x,t) => ListAssoc.memberKey x omega') omega
    in
      ((Utils.listEquals j (List.map #1 subst))
           (* No alpha-equivalence ! Variables must be identical *)
      andalso (envEquals out_subst delta))
           (* The instantiated type must be equal to omega *)
      orelse typeFailureInSeq gamma omega (j, omega') reg
          "ProofChecker.checkSequence: T.EMPTY"
    end

  (* Rule T.CST *)
  | Cst (id, typ, exp, seq) =>
    ((checkExp gamma omega typ reg exp)
    andalso (checkSequence (gamma +! (id, typ)) omega (j, omega') reg seq))
    orelse typeFailureInSeq gamma omega (j, omega') reg
        "ProofChecker.checkSequence: T.CST"

  (* Rule T.VAR *)
  | Var (id, typ, exp, seq) =>
    ((checkExp gamma omega typ reg exp)
    andalso (checkSequence gamma (omega +! (id, typ)) (j, omega') reg seq))
    orelse typeFailureInSeq gamma omega (j, omega') reg
        "ProofChecker.checkSequence: T.VAR "

  (* Rule T.BLOCK *)
  | Comm (Block (seq1, (k, x_tau_list)), seq2) =>
    (case split omega (dom x_tau_list) of
        (* We try to split omega according to the out list *)
      NONE => false (* Failure *)
    | SOME (x_sigma_list, omega'') =>
    (checkSequence gamma x_sigma_list (k, x_tau_list) reg seq1)
    andalso (List.all (fn id => not (Utils.listMember id ((fv_formulas (img gamma))
                                 (* k not-in FV(G,W) *)
                                @ (fv_formulas (img omega))))
                  andalso ((Utils.listMember id j) (* k in j or k not-in FV(W')*)
                  orelse (not (Utils.listMember id (fv_formulas (img omega'))))))
            k)
    andalso (checkSequence gamma (omega'' ++ x_tau_list) (j, omega') reg seq2))
    orelse typeFailureInSeq gamma omega (j, omega') reg
        "ProofChecker.checkSequence: T.BLOCK"

  (* Rule T.INC *)
  | Comm (Inc (id, typ), seq) =>
    (case getVarType id omega of
      SOME (TNat (t)) =>
      (typ = TNat (t)) andalso
      checkSequence gamma (omega +! (id, TNat (s(t)))) (j, omega') reg seq
    | _ => false)
    orelse typeFailureInSeq gamma omega (j, omega') reg
        "ProofChecker.checkSequence: T.INC"

  (* Rule T.DEC *)
  | Comm (Dec (id, typ), seq) =>
    (case getVarType id omega of
      SOME (TNat (t)) =>
      (typ = TNat (t)) andalso
      checkSequence gamma (omega +! (id, TNat (p(t)))) (j, omega') reg seq
    | _ => false)
    orelse typeFailureInSeq gamma omega (j, omega') reg
        "ProofChecker.checkSequence: T.DEC"

  (* Rule T.ASSIGN *)
  | Comm (Affect (id, typ, exp), seq) =>
    (case getVarType id omega of
      NONE => false
    | _ => (checkExp gamma omega typ reg exp)
        andalso (checkSequence gamma (omega +! (id, typ)) (j, omega') reg seq))
    orelse typeFailureInSeq gamma omega (j, omega') reg
        "ProofChecker.checkSequence: T.ASSIGN"

  (* Rule T.FOR *)
  | Comm (For (id, lid, exp, typ, (seq1, (k, sigma_i))), seq2) =>
    (case (typ, split omega (dom sigma_i)) of
        (* We try to split omega according to the out list *)
      (TNat (n), SOME (sigma_0, omega'')) =>
      (envEquals sigma_0 (substituteAll lid FZero sigma_i))
           (* sigma[0] must be obtained *)
      andalso (checkExp gamma omega typ reg exp)
      andalso (List.length k = 0)
           (* k must be empty: no existential quantification in recursion *)
      andalso (checkSequence (gamma +! (id, TNat (FId (lid)))) sigma_i
           (* sigma[i] |- sigma[s(i)] *)
            ([], (substituteAll lid (s(FId (lid))) sigma_i)) reg seq1)
      andalso (checkSequence gamma (omega'' ++ (substituteAll lid n sigma_i))
           (* sigma[n] *)
            (j, omega') reg seq2)
      andalso not (Utils.listMember lid (fv_formulas (img gamma)))
           (* i not_in FV(G) *)
    | _ => false)
    orelse typeFailureInSeq gamma omega (j, omega') reg
        "ProofChecker.checkSequence: T.FOR"

  (* Rule T.CALL *)
  | Comm (ProcCall (exp, typ, exp_typ_list, subst, id_typ_list), seq) =>
    ((checkExp gamma omega typ reg exp)
    andalso List.all (fn (e, t) => checkExp gamma omega t reg e) exp_typ_list
    (* We split both omega and omega' to check type equalities *)
    andalso
      (case (typ, split omega (dom id_typ_list)) of
        (TProc (i, tau_list, k, sigma_list), SOME (_, omega'')) =>
        (Utils.listEquals i (List.map #1 subst))
            (* No alpha-equivalence ! Variables must be identical *)
        andalso
        (let val tau_subst = List.map (substituteList subst) tau_list (* tau[n/i] *)
           val typ_list = List.map #2 exp_typ_list
         in (* exp_list types = tau[n/i] *)
           (ListPair.all (fn (t1, t2) => alphaEqual(t1, t2)) (tau_subst, typ_list))
           orelse typeFailureInSeq (List.map (fn t => (("_", 0), t)) tau_subst)
               (List.map (fn t => (("_", 0), t)) typ_list)
               (j, omega') reg "ProofChecker.checkSequence: T.CALL !!!"
         end)
        andalso (List.length id_typ_list = List.length sigma_list)
        andalso (ListPair.all (fn (t1, t2) => (substituteList subst t1) = t2)
              (sigma_list, List.map #2 id_typ_list)) (* id_list types = sigma[n/i] *)
        andalso (checkSequence gamma (omega'' ++ id_typ_list) (j, omega') reg seq)
        andalso (List.all (fn id => not (Utils.listMember id ((fv_formulas (img gamma))
                                      (* k not-in FV(G,W) *)
                                    @ (fv_formulas (img omega))))
                      andalso ((Utils.listMember id j) (* k in j or k not-in FV(W') *)
                      orelse (not (Utils.listMember id (fv_formulas (img omega'))))))
                k)
      | _ => false))
    orelse typeFailureInSeq gamma omega (j, omega') reg
        "ProofChecker.checkSequence: T.CALL"

  (* Rule T.SUBST-II *)
  | SSubst (seq, delta_i, lid, exp, typ) =>
    ((checkExp gamma omega typ reg exp)
    andalso (case typ of
          TEqual (n, m) =>
          (checkSequence gamma omega (j, (substituteAll lid n delta_i)) reg seq)
                (* delta[n/i] *)
          andalso (envEquals (substituteAll lid m delta_i) omega') (* delta[m/i] *)
        | _ => false))
    orelse typeFailureInSeq gamma omega (j, omega') reg
        "ProofChecker.checkSequence: T.SUBST-II"

  (* Rule T.LABEL *)
  | Comm (Label (id, typ, (seq1, (k, id_typ_list))), seq2) =>
    (case (typ, split omega (dom id_typ_list)) of
      (TProc (k', sigma, [], [TBot]), SOME (x_tau, omega'')) =>
      (Utils.listEquals sigma (img id_typ_list))
      andalso (Utils.listEquals k k')
          (* No alpha-equivalence ! Variables must be identical *)
      andalso (checkSequence (gamma +! (id, typ)) x_tau (k, id_typ_list) reg seq1)
      andalso (checkSequence gamma (omega'' ++ id_typ_list) (j, omega') reg seq2)
    | _ => false)
    orelse typeFailureInSeq gamma omega (j, omega') reg
        "ProofChecker.checkSequence: T.LABEL"

  (* Rule T.JUMP *)
  | Comm (Jump (exp, typ, exp_typ_list, subst, id_typ_list), seq) =>
    ((checkExp gamma omega typ reg exp)
    andalso List.all (fn (e, t) => checkExp gamma omega t reg e) exp_typ_list
    andalso (case (typ, split omega (dom id_typ_list)) of
          (TProc (k, sigma, [], [TBot]), SOME (_, omega'')) =>
          (Utils.listEquals k (List.map #1 subst))
              (* No alpha-equivalence ! Variables must be identical *)
          andalso
            (let val sigma_subst = List.map (substituteList subst) sigma (* sigma[n/i] *)
              val typ_list = List.map #2 exp_typ_list
            in (* We check whether exp_list types = sigma[n/i] *)
              (ListPair.all (fn (t1, t2) => alphaEqual(t1, t2)) (sigma_subst, typ_list))
              orelse typeFailureInSeq (List.map (fn t => (("_", 0), t)) sigma_subst)
                    (List.map (fn t => (("_", 0), t)) typ_list)
                    (j, omega') reg "ProofChecker.checkSequence: T.JUMP !!!"
            end)
          andalso (checkSequence gamma (omega'' ++ id_typ_list) (j, omega') reg seq)
        | _ => false))
    orelse typeFailureInSeq gamma omega (j, omega') reg
        "ProofChecker.checkSequence: T.JUMP"

  | Comm (CommRegion (c, reg), seq) => checkInSequence gamma omega (j, omega')
                                             reg (Comm (c, seq))
  | SeqRegion (seq, reg) => checkInSequence gamma omega (j, omega') reg seq

(**
 * check that the given expression represents a correct proof.
 *
 * @params gamma omega typ reg (expression, tau)
 * @param gamma the constant proof checking environment.
 * @param omega the variable proof checking environment.
 * @param typ the requiered type.
 * @param reg the region.
 * @param expression the expression to check.
 * @param tau the given type.
 *
 * @return true if and only if expression is a correct proof of formula tau.
 *)
and checkExp gamma omega typ reg (expression, tau) =
  (tau = typ) andalso
  case expression of
  (* Rule T.ENV *)
    Id (id) =>
    (case getVarType id (gamma ++ omega) of
      NONE => false
    | SOME t => t = typ)
    orelse typeFailureExp gamma omega typ expression reg
          "ProofChecker.checkExp: T.ENV"

  | Val (v) => checkValue gamma omega typ reg v

  (* Rule T.SUBST-I *)
  | ESubst (exp1, typ1, lid, exp2, typ2) =>
  ((checkExp gamma omega typ2 reg exp2)
  andalso (case typ2 of
        TEqual (n, m) =>
        (checkExp gamma omega (substitute lid n typ1) reg exp1)
        andalso (typ = (substitute lid m typ1))
      | _ => false))
  orelse typeFailureExp gamma omega typ expression reg
      "ProofChecker.checkExp: T.SUBST-I"

  (* Lemma *)
  | Lemma (typlist, typ') =>
  (let val hyp = img (gamma ++ omega)
  in
    (typ = typ') andalso
    (List.all (fn x => Utils.listMember x hyp) typlist)
  end)
  orelse typeFailureExp gamma omega typ expression reg
      "ProofChecker.checkExp: Lemma"

  | ExpRegion (exp, reg) => checkExp gamma omega typ reg (exp, tau)

(**
 * check that the given value represents a correct proof.
 *
 * @params v reg
 * @param v the value.
 *
 * @return true if and only if p is a correct proof.
 *)
and checkValue gamma omega typ reg (value, tau) =
  (tau = typ) andalso
  case value of
  (* Rule T.STAR *)
    Star =>
  (case typ of TEqual (n, m) => true
        | _ => false)
  orelse typeFailureExp gamma omega typ (Val (value, tau)) reg
      "ProofChecker.checkValue: T.STAR"

  (* Rule T.NUM *)
  | Int (n) =>
  (typ = (TNat (mkNat n)))
  orelse typeFailureExp gamma omega typ (Val (value, tau)) reg
      "ProofChecker.checkValue: T.NUM"

  (* Rule T.PROC *)
  | Proc (i, in_list, j, out_list, (seq, (j', typid_list))) =>
  (let val sigmalist = img in_list
     val taulist = img out_list
   in
     (typ = (TProc (i, sigmalist, j, taulist)))
     andalso (Utils.listEquals j j')
         (* No alpha-equivalence ! Variables must be identical *)
     andalso (envEquals out_list typid_list)
     andalso (checkSequence (in_list ++ gamma)
          (List.map (fn (x,t) => (x, top)) typid_list)
          (j', typid_list) reg seq)
     andalso (List.all (fn id => not (Utils.listMember id
                                        (fv_formulas (img gamma)))) i)
                                   (* i not-in FV(G) *)
   end)
  orelse typeFailureExp gamma omega typ (Val (value, tau)) reg
          "ProofChecker.checkValue: T.PROC"

  | ValRegion (v, reg) => checkValue gamma omega typ reg (v, tau)

(**
 * check that the given program represents a correct proof.
 *
 * @params p
 * @param p the program.
 *
 * @return true if and only if p is a correct proof.
 *)
fun checkClosedProg (p : program) =
  case p of
    (SeqRegion (seq, reg), seqType) => checkSequence empty empty seqType reg p
  | _ => raise Fail "ProofChecker.checkClosedProg: not a SeqRegion"
\end{verbatim}

{\newpage}

\section{Source code documentation (Smldoc)\label{sect-smldoc}}

\includegraphics[scale=0.85]{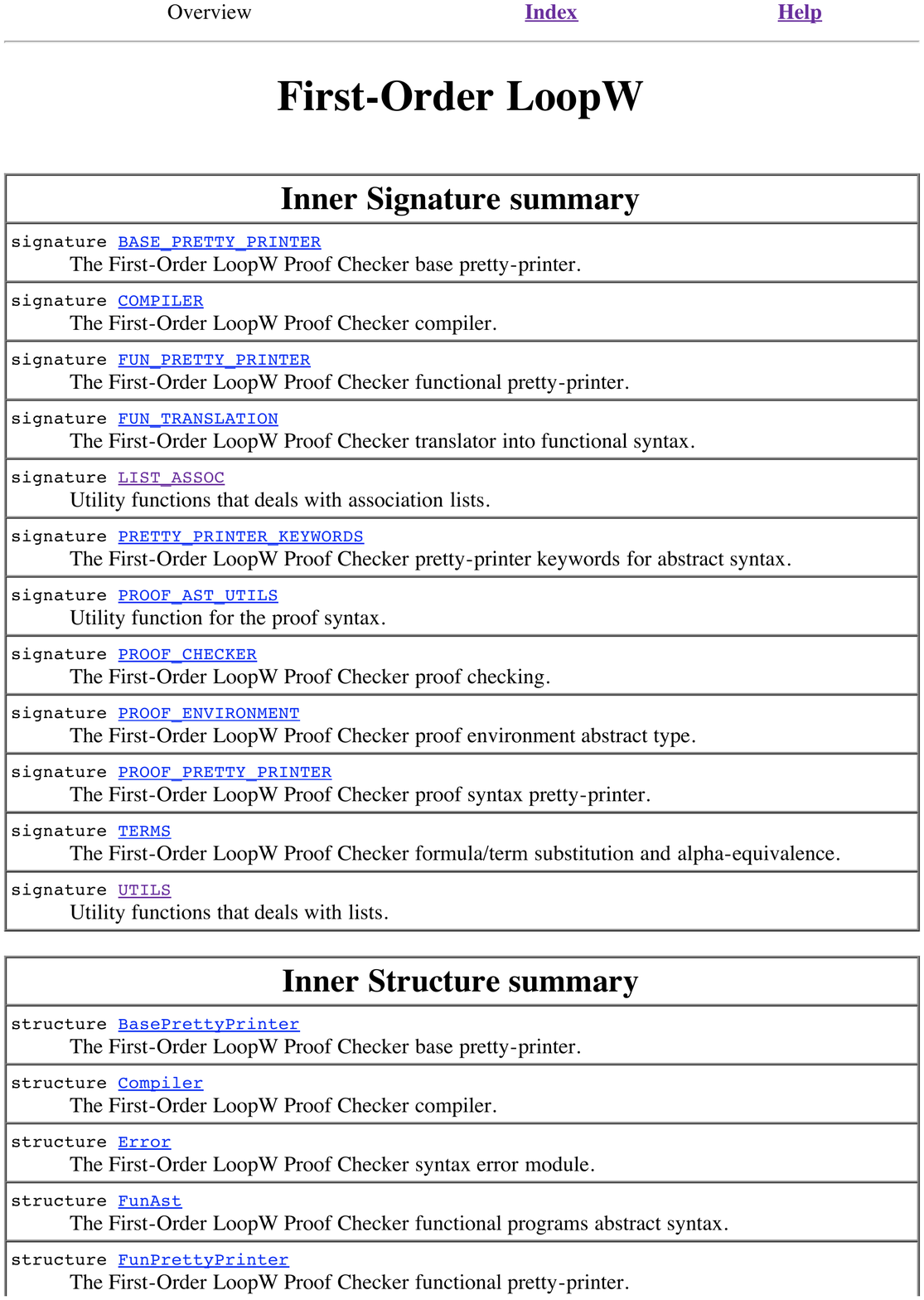}

\includegraphics[scale=0.85]{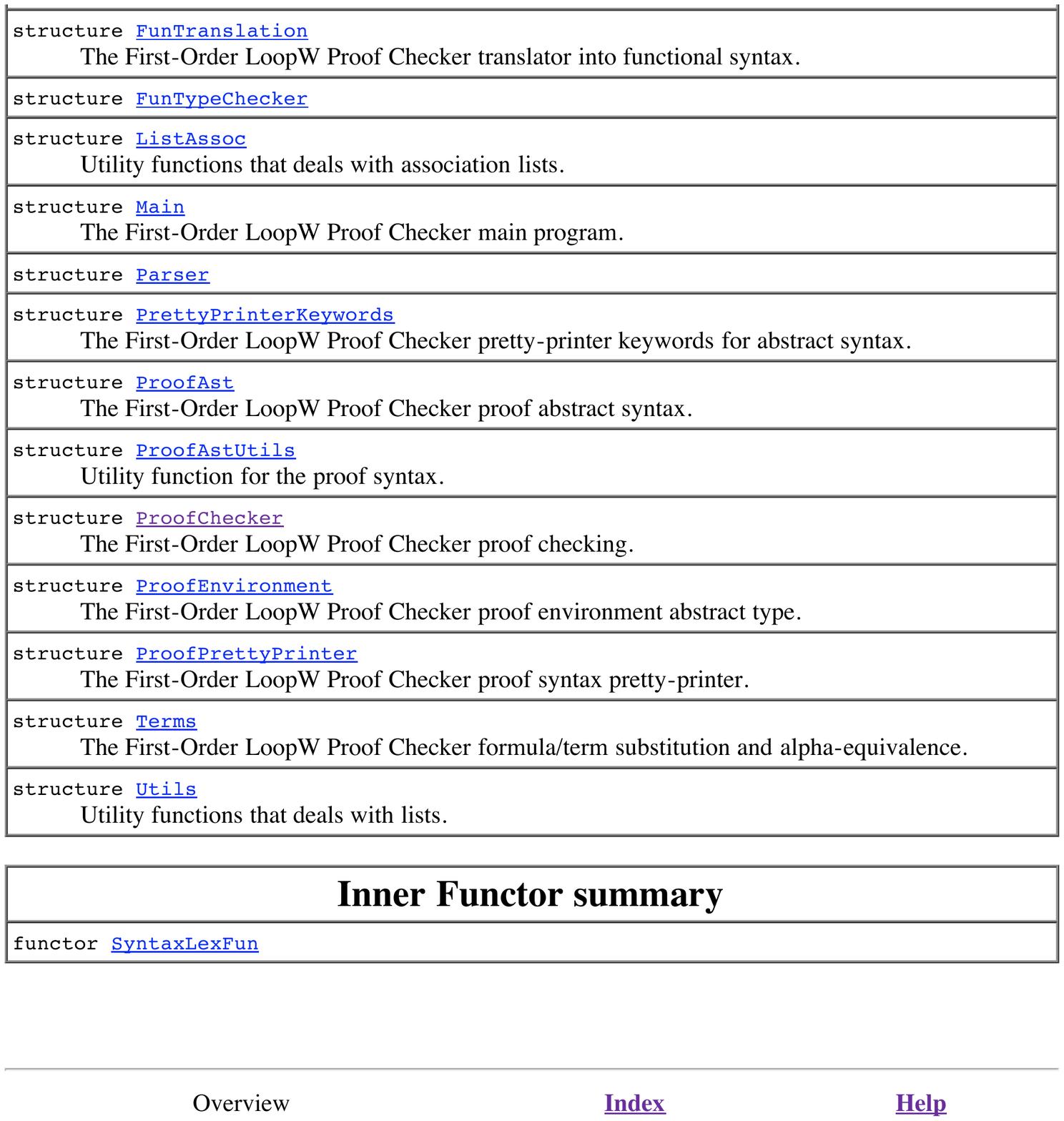}

\includegraphics[scale=0.85]{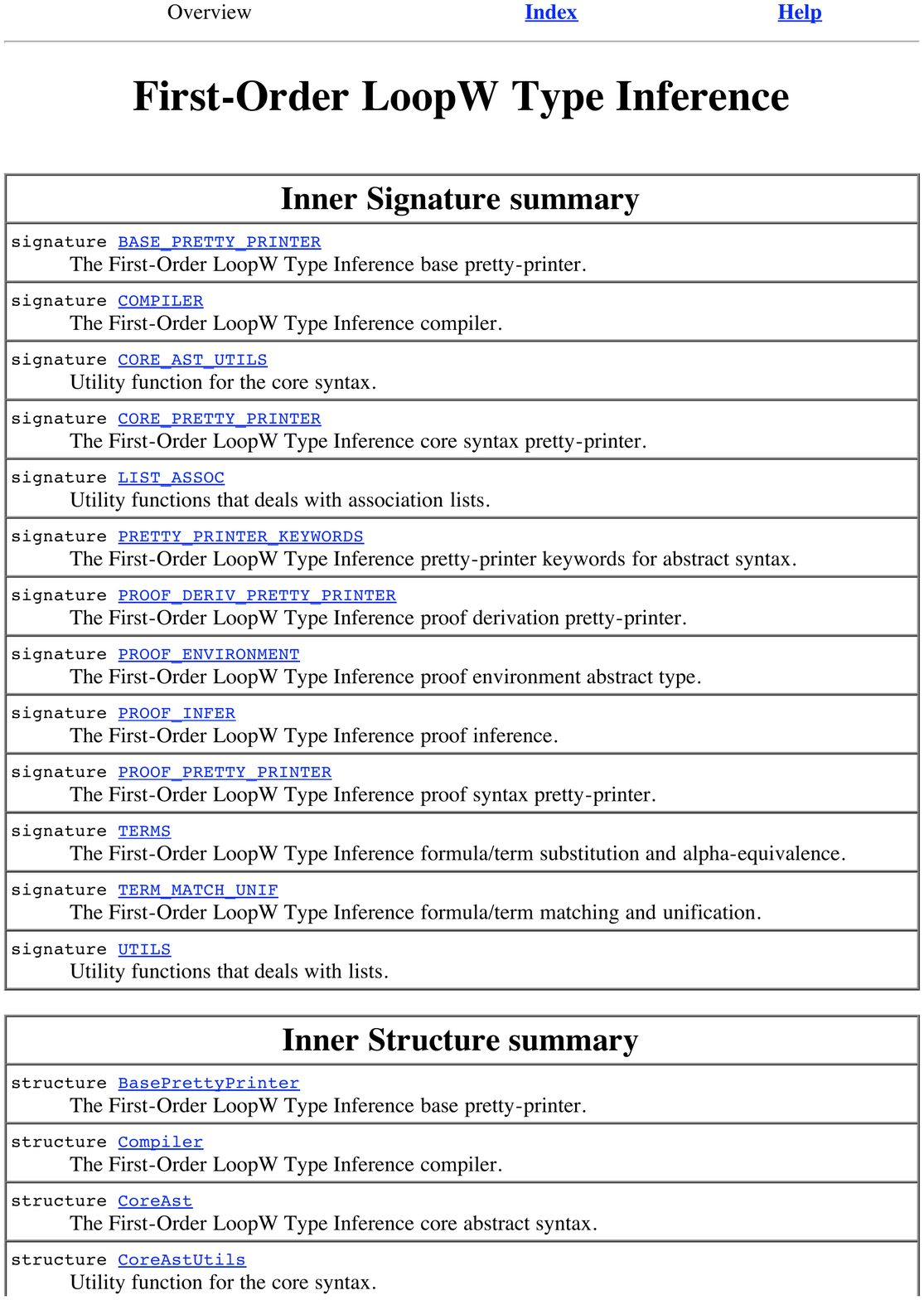}

\includegraphics[scale=0.85]{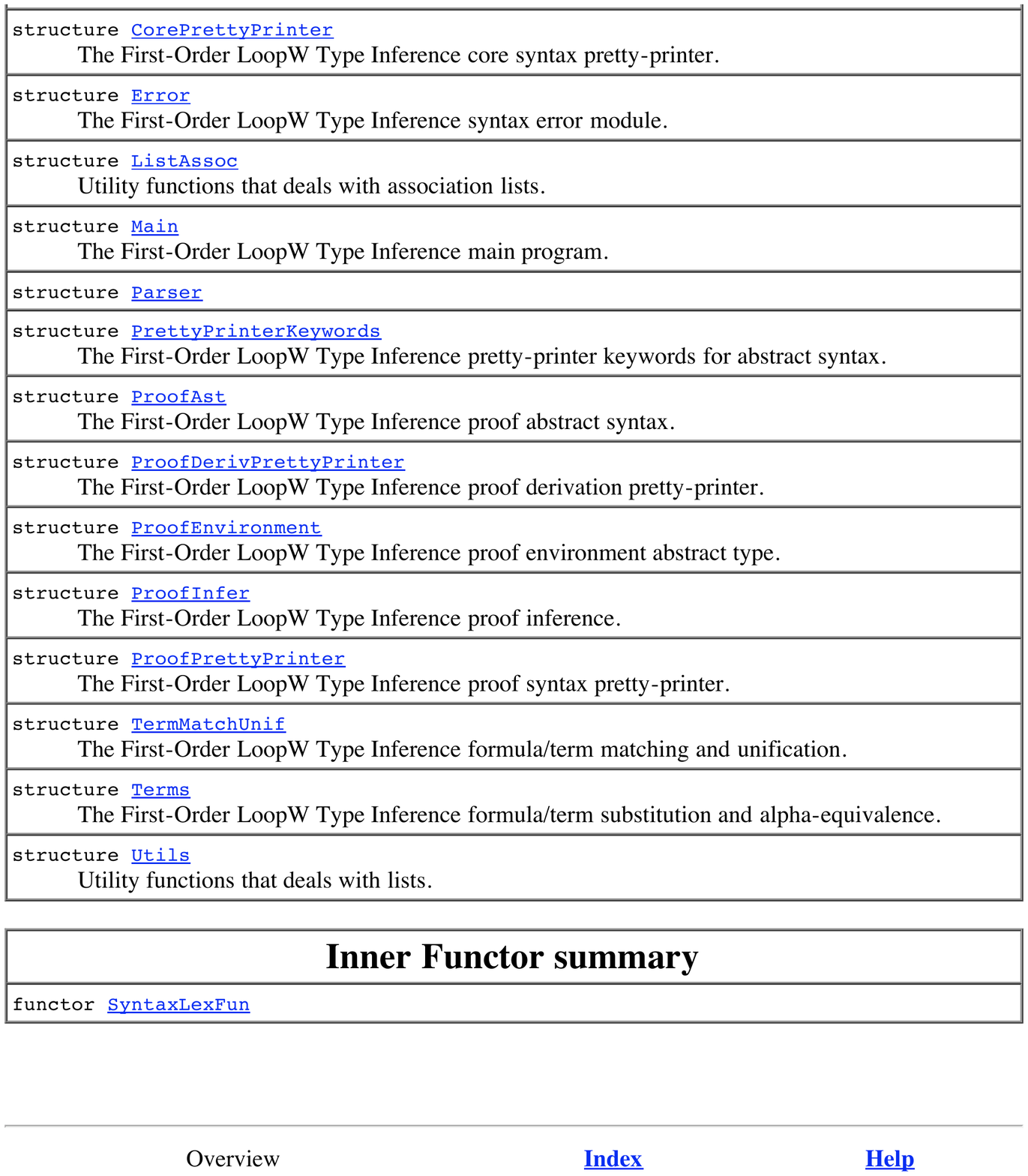}

\end{document}